\documentclass[12pt]{article}
\usepackage{latexsym}
\usepackage{amsmath}
\usepackage{amssymb}
\usepackage{epsfig,graphics}
\usepackage{graphicx}
\usepackage{booktabs}
\usepackage{multirow}
\usepackage{caption}
\usepackage{subcaption}
\usepackage{tikz}
\usepackage{fixmath}
\usetikzlibrary{matrix,snakes,arrows,shapes,decorations.pathmorphing,decorations.markings,calc}

\newcommand{\be}{\begin{equation}}
\newcommand{\ee}{\end{equation}}

\begin{document}

\begin{titlepage}

\vspace*{0.6in}

\begin{center}
{\large\bf On the weak N-dependence of SO(N) and SU(N) \\ gauge theories in 2+1 dimensions}\\
\vspace*{0.75in}
{Andreas Athenodorou$^{a,b}$, Richard Lau$^{c}$ and Michael Teper$^{c}$\\
\vspace*{.25in}
$^{a}$Department of Physics, University of Cyprus, POB 20537, 1678 Nicosia, Cyprus\\
\vspace*{.1in}
$^{b}$Computation-based Science and Technology Research Center, The Cyprus Institute, 20 Kavafi Str., Nicosia 2121, Cyprus \\
\vspace*{.1in}
$^{c}$Rudolf Peierls Centre for Theoretical Physics, University of Oxford,\\
1 Keble Road, Oxford OX1 3NP, UK}
\end{center}

\vspace*{0.4in}

\begin{center}
{\bf Abstract}
\end{center}

We consider (continuum) mass ratios of the lightest `glueballs' as a function of $N$ 
for $SO(N)$ and $SU(N)$ lattice gauge theories in $D=2+1$. We observe that 
the leading large $N$ correction is usually sufficient to describe the 
$N$-dependence of $SO(N\geq 3)$ and $SU(N\geq 2)$, within the errors of the numerical
calculation. Just as interesting is the fact that the coefficient of this correction 
almost invariably turns out to be anomalously small, for both $SO(N)$ and $SU(N)$.
We point out that this can follow naturally from the strong constraints
that one naively expects from the Lie algebra equivalence between certain $SO(N)$
and $SU(N^\prime )$ theories and the equivalence of $SO(\infty)$ and $SU(\infty)$. 
The same argument for a weak $N$-dependence can in principle apply to $SU(N)$ and 
$SO(N)$ gauge theories in $D=3+1$.   

\vspace*{0.95in}

\leftline{{\it E-mail:} athenodorou.andreas@ucy.ac.cy, r.lau1@physics.ox.ac.uk, m.teper1@physics.ox.ac.uk}

\end{titlepage}

\setcounter{page}{1}
\newpage
\pagestyle{plain}

\tableofcontents

\section{Introduction}
\label{section_intro}

Gauge theories at $N\to\infty$ are often more tractable than those at finite $N$
\cite{tHooft_N}.
Since physically relevant theories tend to be at small $N$, it is interesting to 
determine whether the latter are sufficiently `close' to $N=\infty$ for this
limit to be physically useful.

Lattice studies of $SU(N)$ gauge theories in 2+1 and in 3+1 dimensions
do indeed suggest a weak $N$-dependence for the few observables that have been
calculated with adequate precision (see e.g.
\cite{MT98_d3SUN,BLMT02_d3SUN} 
and
\cite{BLMTUW_d4SUN,HMthesis}
respectively),
as do exploratory studies of $SO(N)$ gauge theories in $D=2+1$
\cite{FBRLMT_SON}. 
In this paper we shall use the (preliminary) results of current calculations of glueball 
mass ratios in $SU(N)$
\cite{AAMT_d3SUN} 
and $SO(N)$
\cite{RLMT_d3SON} 
gauge theories in $D=2+1$ to analyse the $N$-dependence with greater reliability and 
accuracy than hitherto. 

We shall find that the $N$-dependence is remarkably weak. (Something that was already 
apparent for $SU(N)$ from earlier calculations.) Not only can the variation
of many mass ratios be accurately described with just a leading $O(1/N^2)$ correction
for $SU(N)$, and $O(1/N)$ for $SO(N)$, but the coefficient of the correction term
turns out to be $\ll 1$. We point out that this can follow naturally
from the strong constraints imposed by the fact that $SO(N)$ and $SU(N)$ gauge
theories share identical $N\to\infty$ planar limits plus the equivalence
between certain $SO(N)$ and $SU(N^\prime)$ theories at smaller $N$.

In Section~\ref{section_prelim} we summarise some expectations for $SO(N)$ and $SU(N)$
gauge theories. We then present in Section~\ref{section_lattice} some results from 
\cite{AAMT_d3SUN,RLMT_d3SON} 
for the $N$-dependence of continuum mass ratios in $SO(N)$ and $SU(N)$ gauge theories. 
The range of $N$ extends up to $N=16$ in both cases, so it is plausible that it 
makes sense to apply large-$N$ expansions. In Section~\ref{section_constraints}
we give examples of some constraints on the $N$-dependence which plausibly arise 
from the (Lie algebra) equivalence between some of the theories. We then briefly
comment on $3+1$ dimensions, and finish with some conclusions.

\section{Preliminaries}
\label{section_prelim}

\subsection{Large $N$}
\label{section_largeN}

In $SU(N)$ gauge theories all-order diagrammatic arguments
\cite{tHooft_N},
supported by non-perturbative lattice calculations (see 
\cite{lattice_revN}
for recent reviews),
suggest that a mass ratio will approach its $N=\infty$ value as
\begin{equation}
\frac{M_i}{M_j}
\stackrel{N\to\infty}{=}
{\tilde{r}}_{ij}
+\frac{\tilde{c}_{1,ij}}{N^2}
+\frac{\tilde{c}_{2,ij}}{N^4}
+ ... \qquad : \quad SU(N) .
\label{eqn_suN}
\end{equation}
In $SO(N)$ gauge theories a similar diagrammatic analysis
\cite{CL_SON}
suggests
\begin{equation}
\frac{M_i}{M_j}
\stackrel{N\to\infty}{=}
{{r}}_{ij}
+\frac{{c}_{1,ij}}{N}
+\frac{{c}_{2,ij}}{N^2}
+ ... \qquad : \quad SO(N)
\label{eqn_soN}
\end{equation}
One can show that the leading  planar diagrams are the same 
in both cases up to a factor of 2 in $g^2$
\cite{CL_SON}.
Moreover $SU(N)$ and $SO(2N)$ gauge theories are related by an
orbifold projection
\cite{orbifold},
and it can be shown that this implies an identical common particle 
spectrum at $N=\infty$
\cite{KUY_orb}. 
So we can expect identical mass spectra at $N=\infty$, i.e.
\begin{equation}
{\tilde{r}}_{ij} = {{r}}_{ij}
\label{eqn_soNsuN}
\end{equation}
in the common $C=+$ sector of the two theories.

\subsection{Small $N$}
\label{section_smallN}

Certain low $N$ pairs of $SO(N)$ and $SU(N^\prime)$ theories are known
to possess the same Lie algebras. These are:  
$SU(2)$ and $SO(3)$, $SU(2)\times SU(2)$ and $SO(4)$, $SU(4)$ and $SO(6)$.
The Lie algebra equivalence suggests that ratios of 
glueball masses may well be identical within each pair of such theories,
in which case
\begin{equation}
\left.\frac{M_i}{M_j}\right|_{SO(3)}=
\left.\frac{M_i}{M_j}\right|_{SU(2)}=
\left.\frac{M_i}{M_j}\right|_{SU(2)\times SU(2)}=
\left.\frac{M_i}{M_j}\right|_{SO(4)}.
\label{eqn_equiv_su2}
\end{equation}
(The single particle spectrum of $SU(2)\times SU(2)$ should be the same 
as that of $SU(2)$, although the former will have extra multiparticle glueball
states consisting of glueballs from the two groups.) We also may expect
\begin{equation}
\left.\frac{M_i}{M_j}\right|_{SO(6)}
=
\left.\frac{M_i}{M_j}\right|_{SU(4)} .
\label{eqn_equiv_su4}
\end{equation}
All this assumes that the differing global structure of the groups does not 
affect the particle spectrum. Whether this plausible assumption is indeed the 
case, or if not whether it is true for some states, is one of 
the interesting questions motivating the $SO(N)$ study in
\cite{RLMT_d3SON}.
It would also be interesting to understand the place of the `Pfaffian' particles
of $SO(2N)$ in this context (see
\cite{Witten98}
for a discussion). 

Note that, to include the $SO(N)$ fundamental string tension in these relations, 
one must take care to match with the correct $SU(N)$ representation. For example, 
an $SO(3)$ confining flux tube carrying fundamental flux corresponds to the 
$SU(2)$ flux tube carrying adjoint flux 
\cite{FBRLMT_SON,RLMT_d3SON}.
In a finite volume some `glueball' states are composed of a 
pair of (conjugate) flux tubes closed around a spatial torus.
Of course as the volume increases these states become heavier
and, eventually, unimportant.

\section{Lattice results}
\label{section_lattice}

The $SO(N)$ results we use are taken from
\cite{RLMT_d3SON}
and those for $SU(N)$ from
\cite{AAMT_d3SUN}.
We refer to these papers for all the details of the calculations.
The methods are entirely standard. The lattice action is the
simple plaquette action. The Monte Carlo is a heat bath for $SO(N)$
\cite{FBRLMT_SON}
and a mixed heat bath plus over-relaxation for $SU(N)$. We use 
a moderately large basis of operators with various spin (J),
parity (P) and charge conjugation (C) quantum numbers and calculate 
their correlators. Their exponential decay as the time separation  
increases provides an estimate of the ground state mass for the specified
$J^{PC}$ quantum numbers. The large basis of operators allows us to
perform a systematic variational calculation which provides
estimates of excited states as well. (Note that we label states 
with the lowest continuum spin $J$ that contributes to the particular 
representation for a square spatial lattice. This will not always 
be correct
\cite{lattice_J}.)

The calculations in 
\cite{RLMT_d3SON}
and in
\cite{AAMT_d3SUN}
attempt to calculate the masses of a large number of excited states.
There are important systematic errors in such calculations, as discussed in
 \cite{RLMT_d3SON,AAMT_d3SUN}.
Here we wish to focus on the typical $N$-dependence of mass ratios 
and so we restrict ourselves to a few of the best determined masses.
We therefore include only the ground states of each (square) representation 
and the first excitations of the lightest of these, i.e. 
the $0^{\pm +}, 2^{\pm +}, 1^{\pm +}$ ground states and the $0^{++\star}$ 
and the $2^{\pm + \star}$ first excited states. 
The lattice volumes used are such that there 
should be no significant contamination from winding flux tube states,
and the masses are not so large that we need to be concerned with 
contamination by multi-glueball states (even at smaller $N$).
However the heavier a state the more rapidly the exponentially
decreasing correlator disappears into the statistical fluctuations 
and (potentially) the larger the systematic error in extracting the mass. 
This is a caveat to consider in the case of the heavier glueballs
i.e. the $0^{-+}$, the $1^{\pm +}$ and the $2^{\pm +\star}$, and particularly
so at coarser lattice spacings. This problem
is enhanced if the overlap of the desired state onto the basis of operators
is smaller -- and this tends to be the case for $SO(N)$ at small $N$.
Nonetheless the systematic error induced by these factors in the qualitative
behaviour of the $N$-dependence of mass ratios -- our main interest here -- 
should not be substantial.

\subsection{$SU(N)$}
\label{subsection_suN}

We calculate 
\cite{AAMT_d3SUN}
our mass ratios in units of the mass gap, the $0^{++}$ ground
state, since this is our most accurately calculated mass. For each $SU(N)$ 
we extrapolate the lattice values of each mass ratio to the continuum limit 
using an $O(a^2)$ correction. We do so for each of  $N=2,3,4,6,8,12,16$. The
resulting mass ratios are plotted in Fig.~\ref{fig_Mm_SUN}. We also plot
best fits of the form 
\begin{equation}
\frac{M}{M_{0^{++}}}
= c_0 + \frac{c_1}{N^2}.
\label{eqn_suNfit}
\end{equation}
The values of $c_0$ and $c_1$ are listed in Table~\ref{table_mMsuNsoN_c1}.

We observe in Fig.~\ref{fig_Mm_SUN} the parity doubling of $J\neq 0$ states
which is expected in the continuum limit of $D=2+1$. (For $J=2$ the parity 
doubling may be broken by finite volume effects, but these should be small 
here.) 

We also observe that the mass ratios can be described with just a leading
$O(1/N^2)$ correction all the way from $SU(16)$ down to $SU(2)$. (We assume
$N=16$ is large enough that there will be no surprises at larger $N$.) 
The $\chi^2$ of these fits is reasonable in most cases. Only for the
$1^{++}$ is it very large ($\sim 6$ per degree of freedom), and for 
the $1^{-+}$ and $0^{++\star}$ it is moderately large 
($\sim 2.5$ per degree of freedom). In the case of $1^{\pm +}$ the
problem is a large (and presumably unphysical)  mass splitting for $SU(6)$, 
rather than the expected
degeneracy, and for the $0^{++\star}$ the problem is a large downward 
fluctuation in $SU(3)$. None of this is helped by including an extra 
$O(1/N^4)$ term in the fit.

The fact that an $O(1/N^2)$ correction suffices for all $N$ already 
tells us that the deviations from $N=\infty$ cannot be large at any $N$.
However a glance at  Fig.~\ref{fig_Mm_SUN} tells us that the deviations
are even smaller than this would suggest if we had natural coefficients 
$c_1 \sim c_0$. Indeed one finds $c_1/c_0 \ll 1$ for the best fits, as 
shown in Table~\ref{table_mMsuNsoN_c1}. One might wonder if this
result is stable under the inclusion of an additional $O(1/N^4)$
correction term (even if the statistical analysis does not demand such 
an extra term). As we see from Table~\ref{table_mMsuNsoN_c1_c2} this 
result is indeed stable. At least in $SU(N)$ glueball mass ratios show
remarkably little variation as $N$ varies from $N=2$ to $N=\infty$.

\subsection{$SO(N)$}
\label{subsection_soN}

We calculate 
\cite{RLMT_d3SON}
the continuum mass ratios in $SO(N)$ just as for $SU(N)$. We do 
so for each of  $N=3,4,5,6,7,8,12,16$. The resulting mass ratios are 
plotted in Fig.~\ref{fig_Mm_SON}. We also plot there
best fits of the form 
\begin{equation}
\frac{M}{M_{0^{++}}}
= c_0 + \frac{c_1}{N}
\label{eqn_soNfit}
\end{equation}
and list the values of $c_0$ and $c_1$ in Table~\ref{table_mMsuNsoN_c1}.

We observe in Fig.~\ref{fig_Mm_SON} the parity doubling of $J\neq 0$ states
just as we saw for $SU(N)$. However it is clear from the scatter of points
that our $SO(N)$ results are considerably less accurate than those for 
$SU(N)$. This must be partly due to the fact that the overlap of the states on
our basis is significantly smaller in $SO(N)$ than in $SU(N)$, particularly
at small $N$, and it may also be that the lack of over-relaxation
in the Monte Carlo update means that it explores the phase space 
more slowly. Nonetheless, the weakness of the $N$-dependence
is evident and we observe that the mass ratios can be described with just 
a leading $O(1/N)$ correction all the way from $SO(16)$ down to $SO(3)$. 
The $\chi^2$ of these fits is reasonable in most cases, only being 
somewhat large, with $\chi^2/n_{dof} \sim 2 -3$, for the $1^{\pm}$ and the 
$2^{\pm \star}$. Fits including an extra $O(1/N^2)$ correction term
improve the $\chi^2/n_{dof}$ for the $2^{+\star}$ and the $1^+$ but not 
for the  $2^{-\star}$ and the $1^-$. Since the $P=\pm$ pairs of states 
should be degenerate (at each $N$) it is not clear if the need for such 
an extra term in these states is being indicated or not.  

In any case, what we clearly see is a very weak $N$-dependence
that in most cases can be described with a leading $O(1/N)$ correction.
Moreover, just as for $SU(N)$, the coefficient of this correction is 
small, $c_1/c_0 \ll 1$, as we see in Table~\ref{table_mMsuNsoN_c1}. 
One might again wonder if this result is stable under the inclusion 
of an additional $O(1/N^2)$ correction term and the indications from 
Table~\ref{table_mMsuNsoN_c1_c2} are that this is indeed the case,
albeit with large uncertainties. So, just as for $SU(N)$, the glueball 
mass ratios in $SO(N)$ show remarkably little variation with $N$
over the whole range of $N$.

\subsection{$SO(N)$ and $SU(N)$: a comparison}
\label{subsection_compare}

We see from the best fits listed in Table~\ref{table_mMsuNsoN_c1} 
that the $N\to\infty$ limits of the $SU(N)$ and $SO(N)$ mass ratios
are very similar and, given that the errors listed are purely
statistical, that they are broadly compatible. The apparent differences 
are $\sim 1 - 5\%$ with the $SO(N)$ values being always higher. This is 
in the direction one would expect from the smaller $SO(N)$ overlaps 
leading to a slightly too-early identification of the effective mass 
plateaux that then leads to a small systematic over-estimate of the masses.

There is no obvious best way to compare the  $SU(N)$ and $SO(N)$ 
mass ratios  at finite $N$ given the different powers of the
leading corrections. Here we shall simply overlay in
Figs.~\ref{fig_Mm_suNsoNa},\ref{fig_Mm_suNsoNb} the mass ratios 
for $SU(N)$ and $SO(N)$. (The reader can use the fits in 
Table~\ref{table_mMsuNsoN_c1} to construct alternative comparisons.) 
Without going into fine details (see
\cite{RLMT_d3SON}
for a careful comparison) we see that the mass ratios of the
two theories are broadly similar, taking into account the larger
systematic errors on the most massive states. A similar comment applies
to the comparison between $SU(2)$ and $SO(3)$, $SU(2)$ and $SO(4)$, and 
$SU(4)$ and $SO(6)$, with similar caveats concerning the
most massive states.

The reader will have noticed that so far we have not considered
the fundamental string tension, $\sigma_f$, in our mass ratios. This
is usually the physical quantity that is most accurately obtained in
lattice calculations of energies and so one often sees continuum 
glueball masses presented as a ratio $M/\surd\sigma$. The reason we 
have not done so is that, for example, the fundamental 
$f={\underline{3}}$ of $SO(3)$ corresponds to the adjoint 
$A={\underline{3}}$ of $SU(2)$, and the $f={\underline{6}}$ 
of $SO(6)$ corresponds to the $k=2$ antisymmetric of $SU(4)$.
That is to say, the fundamental string tensions for e.g. $SO(3)$
and $SU(2)$ are not the same physical quantities. This is in contrast
to colour singlet glueball masses that do not care about the
representation of the fundamental fields. 
Since $\sigma_A[SU2] \sim 2.5 \sigma_f[SU2]$
\cite{AAMT_d3SUN,RLMT_d3SON}
and  $\sigma_{2a}[SU4] \sim 1.35 \sigma_f[SU4]$ 
(see e.g.
\cite{string_rep})
we know in advance that a mass ratio $M/\surd\sigma_f$ will have
a strong $N$-dependence for either $SU(N)$ or for $SO(N)$
or for both. To illustrate this we show in Fig.~\ref{fig_MK_suNsoN}
the ratio of the mass of the ground state $J^{PC}=0^{++}$ glueball (the
mass gap) to the fundamental string tension for both $SO(N)$ and $SU(N)$.
The corrections to the $N=\infty$ limit are clearly much greater
than in the mass ratios shown in  Figs.~\ref{fig_Mm_SUN},~\ref{fig_Mm_SON}.
It is equally clear from  Fig.~\ref{fig_MK_suNsoN} 
that a straight line fit to the $SO(N\geq 3)$ 
ratios will not work: one needs to include an $O(1/N^2)$ 
term in addition to the leading $O(1/N)$ term. This is also
the case for the $SU(N\geq 2)$ ratios: one needs to include an
$O(1/N^4)$ term in addition to the leading $O(1/N^2)$ term.

\section{Constraints on the N-dependence}
\label{section_constraints}

We assume in this section that the glueball spectrum of both $SO(3)$
and $SO(4)$ is the same as that of $SU(2)$ and that the spectrum of
$SO(6)$ is the same as that of $SU(4)$. (In the case of $SO(4)$ 
there will be extra multi-glueball states from the $SU(2)\times SU(2)$
structure, but that does not affect our argument here.) We also
assume the $N=\infty$ glueball spectra of $SO(N)$ and $SU(N)$ are 
identical in their common $C=+$ sector. These constraints become
quite powerful when we assume in addition that we only need a few of 
the terms in the expansions in eqns(\ref{eqn_suN},\ref{eqn_soN}) to 
describe the glueball spectra $\forall N$. This last assumption is 
quite strongly supported by the lattice calculations which, as we have
seen, typically require only a leading order correction to reproduce the
spectra for all $N$, within the errors.

How strongly these constraints determine the mass spectra will depend 
on how many terms we need to retain in the expansions in 
eqns(\ref{eqn_suN},\ref{eqn_soN}) to accurately reproduce the
mass spectra for all $N$. We illustrate the possibilities
with the following sample of scenarios.

\vspace*{0.2cm}

1. Assume we know the $SU(N)$ spectrum. Then, given our above
assumption, we also know the
spectra of $SO(3)$, $SO(4)$, $SO(6)$ and $SO(\infty )$. This is enough
to predict the spectrum of $SO(N)$ for all $N$, if the large $N$
expansion, when truncated to 4 terms,
\begin{equation}
\frac{M}{M_{0^+}}
\stackrel{N\geq 3}{\simeq}=
c_0
+\frac{{c}_{1}}{N}
+\frac{{c}_{2}}{N^2}
+\frac{{c}_{3}}{N^3} \qquad : \quad SO(N),
\label{eqn_soNb}
\end{equation}
is sufficiently accurate, as is strongly supported by
our calculations which show that just the leading term is mostly good 
enough within our statistical errors. That is to say: the $SU(N)$
spectrum predicts that of $SO(N)$ $\forall N$ within the accuracy
of eqn(\ref{eqn_soNb}).
An example of such a prediction was displayed in Fig.1 of
\cite{FBRLMT_SON}.

\vspace*{0.2cm}

2. Suppose that the spectrum is accurately reproduced by
eqn(\ref{eqn_soNb}) with only the first $O(1/N)$ correction
non-zero. Then the equality of the $SO(3)$ and $SO(4)$ spectra, 
immediately tells us that $c_1=0$, i.e. there is no $N$-dependence at all 
in $SO(N)$. This then demands that the $SU(2)$, $SU(4)$ and $SU(\infty)$ 
spectra should also be equal. So if the expansion for $SU(N)$
\begin{equation}
\frac{M}{M_{0^+}}
\stackrel{N\geq 2}{\simeq}=
\tilde{c}_0
+\frac{\tilde{c}_{1}}{N^2}
+\frac{\tilde{c}_{2}}{N^4} + ... \qquad : \quad SU(N),
\label{eqn_suNb}
\end{equation}
is sufficiently accurate with just the first two correction terms (not 
implausible given what the $SU(N)$ lattice calculations indicate) we 
have no $N$-dependence for $SU(N)$ either. 

\vspace*{0.2cm}

3. Suppose that two correction terms, i.e. $c_1/N+c_2/N^2$, suffice for 
$SO(N\geq 3)$. Subtracting the expansions for $SO(3)$ and $SO(4)$, 
and using the equality of the mass ratios, we see that $c_2 = -12c_1/7$. 
This can then be used to reduce the number of fitted parameters 
from two to one. If we additionally assume that a single 
correction term, i.e. $\tilde{c}_1/N^2$, suffices for $SU(N\geq 2)$
(as often appears to be the case) then we have three relations between 
these coefficients (using also $c_0=\tilde{c}_0$) whose only solution is
$\tilde{c}_1 = c_1 = c_2 = 0$, i.e. no $N$-dependence at all.

\vspace*{0.2cm}

4. Suppose we ignore the constraints from $SU(2)$ because, for example,
we do not trust the large-$N$ expansion for such low $N$, and similarly
for $SO(3)$. In such a case, if we make the 
relatively weak assumption that a single correction, $\tilde{c}_1/N^2$, 
suffices for $SU(N\geq 4)$, and similarly a single correction, 
${c}_1/N$, suffices for $SO(N\geq 6)$ then we immediately 
obtain $\tilde{c}_1 = 16c_1/6$, i.e. the $N$-dependence
of $SU(N\geq 4)$ is completely constrained by that of $SO(N\geq 6)$
(or vice versa). 

\vspace*{0.2cm}

An important feature of these arguments is that if the number of
significant terms in expansions around $N=\infty$ is small enough, 
then the expansion coefficients will be zero. Since the lattice 
calculations are indeed consistent with such a small number of terms, 
this perhaps provides an explanation why the coefficients of these 
terms turn out to be unexpectedly small and why there is so little
dependence on $N$ for both the $SO(N)$ and $SU(N)$ glueball spectra.

Given the importance of our lattice results in supporting such 
arguments, and given that our errors are finite, we need to ask how
large a higher order term might be concealed within these errors. To 
address this question we show in Table~\ref{table_mMsuNsoN_c1_c2}
the subleading coefficient, $c_2$, in fits that include such a term.
We do so only for $SU(N)$ since the errors in $SO(N)$ are too large
to provide any kind of tight constraint. What we see is that apart
from $J=1$, where the fits are very poor and the systematic errors
are largest, the coefficients of the $O(1/N^4)$ term are small
and indeed consistent with zero (as one would expect given that
the leading term by itself gives acceptable fits). So for $SU(N)$
at least, arguments based on a low order expansion in $1/N^2$
do appear to have a significant motivation. 

Of course future calculations with much smaller errors will inevitably
expose the presence of higher order terms in the $1/N$ expansions.
Our above analysis suggests that the coefficients of these will be 
small enough for the arguments of this section to retain an approximate
validity. However this will certainly make the comparison between $SO(N)$ 
and $SU(N)$ more delicate, in much the same way as in the comparison 
between $SU(N)$ adjoint and bifundamental chiral condensates 
\cite{adj_bifun} 
where the quantities are very accurately determined and the
corrections are respectively in powers of $1/N^2$ and $1/N$, 
just as in our case. 

\section{D=3+1}
\label{section_d4}

The discussion in Sections~\ref{section_prelim} and \ref{section_constraints}
carries over unchanged to $D=3+1$. So the question is whether lattice calculations
encourage us to assume that low-order large-$N$ expansions are accurate
all the way down to $SU(2)$ and $SO(3)$, or not. 

For $SU(N)$ some calculations exist. In Fig.~\ref{fig_Mm_d4SUN} we show some 
results from 
\cite{HMthesis}
and
\cite{BLMTUW_d4SUN}.
We see that the $2^{++}$ ground and first excited states show little variation 
with $N$. The first excited $0^{++}$ is however not consistent between the
two calculations: one indicates a large variation, the other a modest one!
The main conclusion here must be that much more accurate lattice calculations
are needed if we wish to pursue this question.

In the case of $SO(N)$ a few calculations exist,
\cite{FBRLMT_SON},
but at no $N$ has a continuum extrapolation been performed. The problem here
is that there is a first-order phase transition in the lattice (bare) coupling,
separating the weak and strong coupling phases, which, for small $N$, occurs
for a very small lattice spacing (measured on the weak coupling side)
\cite{FBRLMT_SON}.
Thus for small $N$ extremely large lattice volumes are needed if one is
to be on the weak coupling side, from where one can take a continuum limit.
It may be that improved lattice actions will help to overcome this
obstacle, but at present no calculations useful for our purposes 
exist for $SO(N)$ in $D=3+1$.

\section{Conclusions}
\label{section_concl}

In this paper we have considered the $N$-dependence of glueball masses, 
in units of the mass gap, for $SU(N)$ and $SO(N)$ gauge theories in
$2+1$ dimensions. Out of the glueball spectra, calculated in 
\cite{AAMT_d3SUN,RLMT_d3SON}, 
we have selected a few of the most reliably and precisely 
calculated states. We have seen that the $N$ dependence is very weak and
in most cases can be described with just the leading large-$N$ correction
for all values of $N$. Moreover we saw that the coefficient of the
correction term is unexpectedly small. We noted that the 
Lie algebra equivalences between certain $SO(N)$  and $SU(N^\prime)$
groups can become very constraining if the $N$-dependence of the spectra 
can be described by sufficiently few terms in the large-$N$ expansions
for both $SU(N)$ and $SO(N)$, and that this provides a possible explanation
for the very weak $N$-dependence. In principle such arguments carry over to 
$D=3+1$ but at present we do not have usefully precise indications that
one or two correction terms suffice to describe the $N$-dependence of these
theories.

\section*{Acknowledgements}

AA has been partially supported by an internal program of the University of Cyprus 
under the name of BARYONS. In addition, AA acknowledges the hospitality of the 
Cyprus Institute where part of this work was carried out. 
RL has been supported by an STFC STEP award under grant ST/J500641/1, 
and MT acknowledges partial support under STFC grant ST/L000474/1.
The numerical computations were carried out on EPSRC and Oxford 
University funded computers in Oxford Theoretical Physics.

\clearpage

\begin{table}[htb]
\begin{center}
\begin{tabular}{|c|cc|cc|}\hline
 & \multicolumn{2}{|c}{ $-c_1/c_0$ } & \multicolumn{2}{|c|}{ $c_0$ } \\ \hline
state & $SO(N)$  & $SU(N)$  & $SO(N)$  & $SU(N)$  \\ \hline
$0^{++\star}$   & 0.278(46)  &  0.223(13)  & 1.593(12) & 1.530(3)  \\
$0^{-+}$       & 0.030(70)   &  0.183(21)  & 2.193(27) & 2.183(6)  \\
$2^{++}$      & 0.139(41)   &  0.091(14)  &  1.744(13) & 1.679(3)   \\
$2^{-+}$      & 0.110(42)   &  0.078(14)  & 1.720(14) & 1.681(4)   \\
$2^{++\star}$  & 0.098(61)  &   0.239(17)  & 2.132(23) & 2.048(5) \\
$2^{-+\star}$  & 0.050(58)  &   0.243(18)  & 2.111(22) & 2.055(4) \\
$1^{++}$      & 0.118(75)  &  0.329(21)   & 2.548(36) & 2.427(7) \\ 
$1^{-+}$      & 0.164(86)  &  0.295(22)   & 2.560(40) & 2.407(7) \\ \hline
\end{tabular}
\caption{Coefficient of leading large-$N$ fits  to mass
ratios $M/M_{0^{++}}$ in $SO(N)$ and $SU(N)$, using fits $c_0-c_1/N^2$
for $SU(N)$ and $c_0-c_1/N$ for $SO(N)$. Errors are statistical.}
\label{table_mMsuNsoN_c1}
\end{center}
\end{table}

\begin{table}[htb]
\begin{center}
\begin{tabular}{|c|c|cc|}\hline
\multicolumn{1}{|c}{state} & \multicolumn{1}{|c}{$SO(N)$} & \multicolumn{2}{|c|}{$SU(N)$}  \\ \hline
      &  $|c_1|/c_0$ &  $|c_1|/c_0$ &  $|c_2|/c_0$ \\ \hline
$0^{++\star}$   & 0.47(21)  &     0.266(48)   &  0.16(18) \\
$0^{-+}$       & 0.25(30)   &     0.198(77)  &  0.05(27) \\
$2^{++}$      & 0.17(18)   &     0.117(36)   &  0.10(13) \\
$2^{-+}$      & 0.24(19)   &     0.145(52)   &  0.25(19) \\
$2^{++\star}$  & 0.38(27)  &      0.190(72)  &  0.18(26) \\
$2^{-+\star}$  & 0.04(25)  &      0.249(75)   & 0.03(28) \\
$1^{++}$      & 0.95(38)  &     0.191(77)   &  0.51(27) \\ 
$1^{-+}$      & 0.59(50)  &     0.010(90)   &  1.13(32) \\ \hline
\end{tabular}
\caption{Normalised coefficient of leading (and sub-leading) large-$N$ correction to mass
ratios $M/M_{0^{++}}$ in $SO(N)$ and $SU(N)$ from fits $c_0+c_1/N^2+c_2/N^4$
for $SU(N)$ and $c_0+c_1/N+c_2/N^2$ for $SO(N)$. Errors are statistical.}
\label{table_mMsuNsoN_c1_c2}
\end{center}
\end{table}

\begin{figure}[htb]
\begin	{center}
\leavevmode
\input	{plot_Mm_suN.tex}
\end	{center}
\caption{Some masses versus $1/N^2$ in $SU(N)$, in units of the mass gap. 
In ascending order: the first excited $J^{PC} = 0^{++}$, $\Box$, the $2^{++}, \, \bullet$, 
and $2^{-+}, \, \circ$, ground states, the $2^{++}, \, \bullet$, 
and $2^{-+}, \, \circ$, first excited states, the $0^{-+}$ ground state, $\Box$, and the 
$1^{++}, \, \bullet$, and $1^{-+}, \, \circ$, ground states. The $P=-$ partners
have been shifted horizontally to be more visible. Lines are 
corresponding best fits of the form $c_0 + c_1/N^2$.}
\label{fig_Mm_SUN}
\end{figure}

\begin{figure}[htb]
\begin	{center}
\leavevmode
\input	{plot_Mm_soN.tex}
\end	{center}
\caption{Some masses versus $1/N$ in $SO(N)$, in units of the mass gap. 
In ascending order: the first excited $J^{P} = 0^{+}$, $\Box$, the $2^{+}, \, \bullet$,
and $2^{-}, \, \circ$, ground states, the $2^{+}, \, \bullet$, and $2^{-}, \, \circ$, 
first excited states, the $0^{-}$ ground state, $\Box$, and the 
$1^{+}, \, \bullet$, and $1^{-}, \, \circ$, ground states.  The $P=-$ partners
have been shifted horizontally to be more visible. Lines are 
corresponding best fits of the form $c_0 + c_1/N$.}
\label{fig_Mm_SON}
\end{figure}

\begin{figure}[htb]
\begin	{center}
\leavevmode
\input	{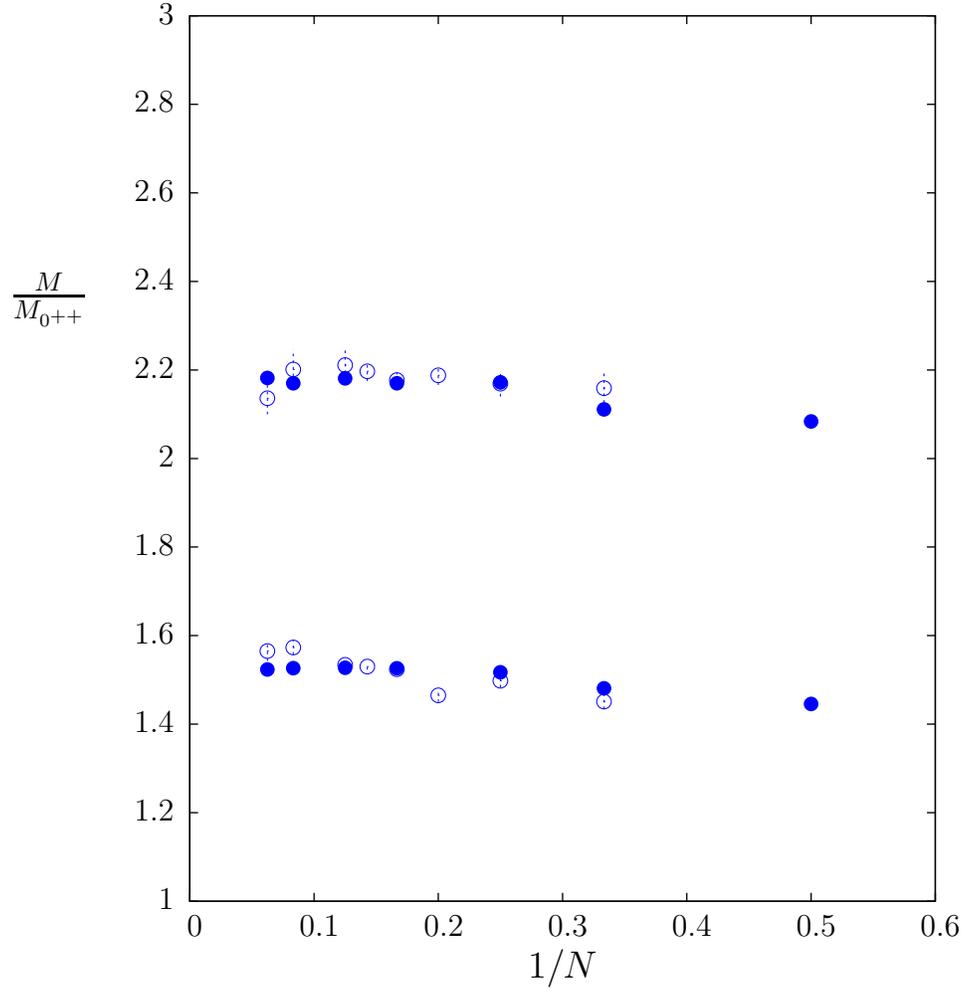}
\end	{center}
\caption{Some masses versus $1/N$ in $SO(N)$, $\circ$, and $SU(N)$, $\bullet$, 
in units of the mass gap. 
In ascending order: the first excited $J^{PC} = 0^{++}$ and the $0^{-+}$ ground state.}
\label{fig_Mm_suNsoNa}
\end{figure}

\begin{figure}[htb]
\begin	{center}
\leavevmode
\input	{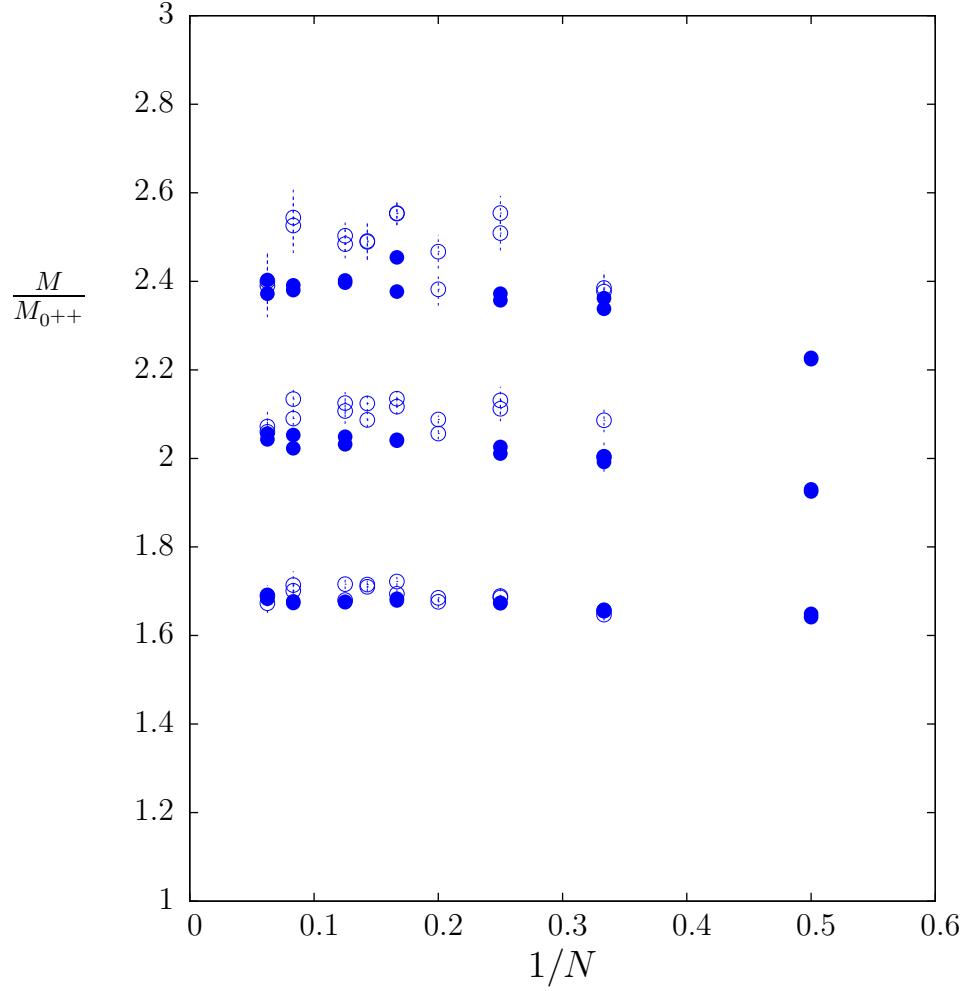}
\end	{center}
\caption{Some masses versus $1/N$ in $SO(N)$, $\circ$, and $SU(N)$, $\bullet$, 
in units of the mass gap. In ascending order:  the $J^{PC} = 2^{\pm +}$ ground states, 
the $2^{\pm +}$ first excited states, and the $1^{\pm +}$ ground states.  }
\label{fig_Mm_suNsoNb}
\end{figure}

\begin{figure}[htb]
\begin	{center}
\leavevmode
\input	{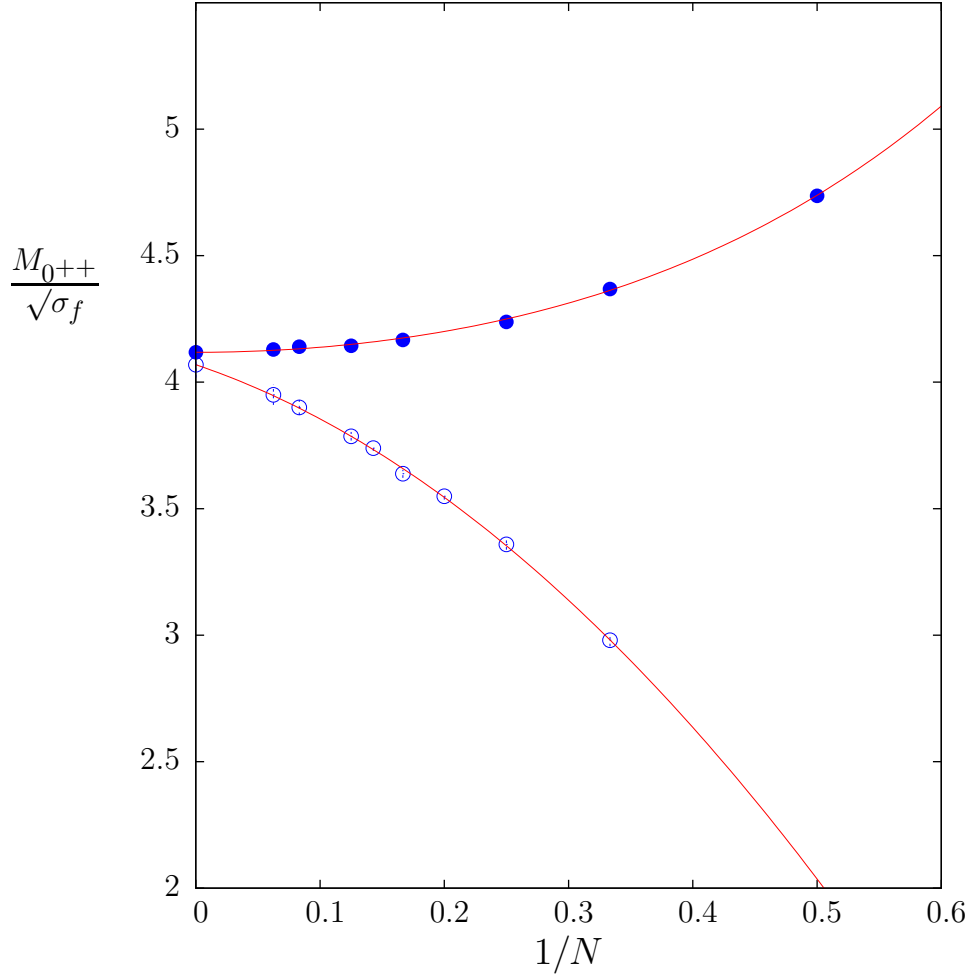}
\end	{center}
\caption{The mass of the lightest scalar glueball in units of the fundamental 
string tension versus $1/N$: for $SO(N)$, $\circ$, and for $SU(N)$, $\bullet$. 
Lines are best fits as described in text. Extreme left points are values extrapolated 
to $N=\infty$.}
\label{fig_MK_suNsoN}
\end{figure}

\begin{figure}[htb]
\begin	{center}
\leavevmode
\input	{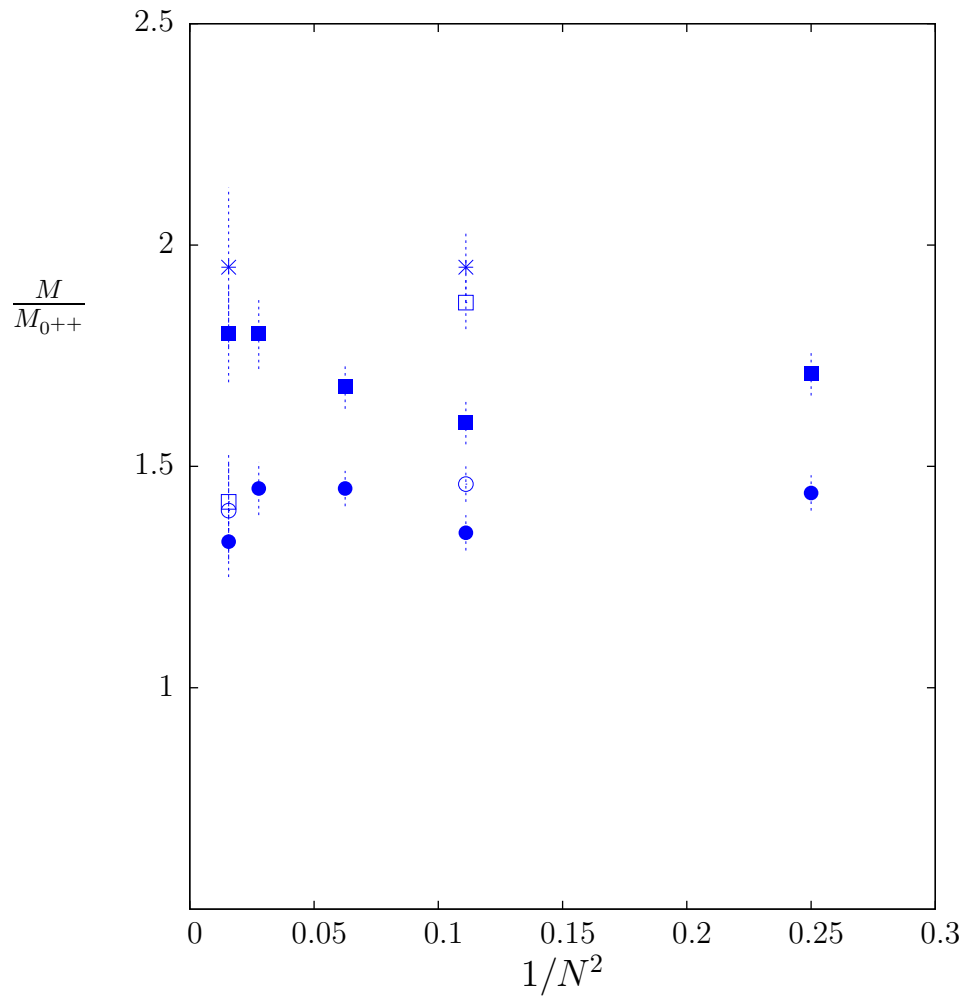}
\end	{center}
\caption{Some masses versus $1/N^2$ in $SU(N)$ in $D=3+1$, in units of the mass gap. 
The first excited $J^{PC} = 0^{++}, \, \square$, the ground state $2^{++}, \, \circ$, 
and the first excited  $2^{++}, \, \star$: all from \cite{HMthesis}.
Also the first excited $J^{PC} = 0^{++}, \, \blacksquare$, and  the $2^{++}, \, \bullet$,  
ground state from \cite{BLMTUW_d4SUN}.}
\label{fig_Mm_d4SUN}
\end{figure}

\end{document}